\documentclass[aps,prc,preprint,groupedaddress]{revtex4-2}

\begin{document}

\title{\bf Generating function for nucleus-nucleus scattering amplitudes
in Glauber theory}

\author{Yu.M. Shabelski}
\email[]{shabelsk@thd.pnpi.spb.ru}
\author{A.G. Shuvaev}
\email[]{shuvaev@thd.pnpi.spb.ru}
\affiliation{Petersburg Nuclear Physics Institute, Kurchatov National
Research Center,
Gatchina, St. Petersburg 188300, Russia
}

\begin{abstract}
A new approach to deal with the scattering
amplitudes in Glauber theory is proposed. It relies on the use
of generating function, which has been explicitly found.
The method is applied to the analytical calculation of
the nucleus-nucleus elastic scattering amplitudes
in the all interaction orders of the Glauber theory.
\end{abstract}

\maketitle

\section{Introduction}
The theory of nucleus-nucleus interaction is an important aspect
of a long-standing multiple scattering problem.
It has acquired a modern impetus from
the large number of the currently available experimental data
(see e.g. Refs.~\cite{Tanihata:1985zq, Tanihata:1986kh, Smith:2008zh,
Bertulani:2009zw}).
The theoretical calculations provide,
in particular, a way to get information on scattering of both stable
and unstable nuclei at the comparatively high energies
of more than several hundreds~of MeV per nucleon.
The calculations are standardly carried out in the Glauber
approach~\cite{Glaub1, Glauber:1970jm}. It has proven to be highly efficient
for the hadron-nucleus collision,
supplying rather simple analytical expressions for the scattering
amplitudes. The case of the nucleus-nucleus scattering is much more
involved. Additional simplifying approximations
are commonly used to obtain an
analytical expression such as the optical model or the rigid target
model (see e.g.Refs.~\cite{Czyz:1969jg, Bialas:1977pd, Alkhazov:1977ur}).
Apart from these models there are only numerical calculations
based on the Monte-Carlo method or on its modifications~\cite{Zadorozhnyi:1983fz,
Gareev, Lobodenko, Merino:2009yj}.

In the present paper we propose a novel approach.
Assuming the range of nucleon-nucleon interaction to be small compared
to the typical nucleus size we have derived the analytical expression
for the generating function giving the Glauber amplitudes
for nucleus-nucleus scattering.
Its relatively simple form allows one to reach
the same accuracy as that provided with the numerical
Monte-Carlo calculations without being as time-consuming as they are.

\section{Scattering amplitudes in the Glauber theory}
To begin, we briefly outline the basics of the Glauber theory.
The amplitude of the elastic scattering of the incident nucleus
$A$ on the fixed target nucleus $B$
reads~\cite{Pajares:1983gw, Braun:1988pk}
\begin{equation}
\label{dcs}
F_{AB}^{el}(q)\,=\,\frac{ik}{2\pi}\int d^{\,2}b\,e^{iqb}
\bigl[1\,-\,S_{AB}(b)\bigr],
\end{equation}
where $q$ is the transferred momentum and $k$ is the mean momentum
carryed by a nucleon in nucleus $A$. The two-dimensional impact
momentum $b$ lies in the transverse plane to the vector $k$.
The main assumption underlying the Glauber theory is
that the radius of the nucleon-nucleon interaction is
much smaller than the typical nucleus size.
Then assuming the phase shifts of the nuclear scattering
to be the sum of those for each nucleon-nucleon scattering,
$\chi_{NN}(b)$, the function $S_{AB}(b)$ takes the form
\begin{equation}
\label{I}
S_{AB}(b)\,=\,\langle\,A,\,B\,|
\left\{\prod\limits_{i\,j}\bigl[1-\Gamma_{NN}(b+x_i-y_j)\bigr]
\right\}
|\,A,\,B\,\rangle,
\end{equation}
with
\begin{equation}
\label{chi}
\Gamma_{NN}(b)\,=\,1\,-\,e^{i\chi_{NN}(b)}\,=\,
\frac 1{2\pi ik}\int d^{\,2}q\,e^{-iqb}f_{NN}^{el}(q),
\end{equation}
where $f_{NN}^{el}(q)$ is the nucleon-nucleon scattering
amplitude.
The product in Eq.~(\ref{I}) comprises all pairwise interactions
between the nucleons from the projectile and target nuclei
$A$ and $B$, with symbol $\langle\,A,\,B\,|\,\cdots\, |\,A,\,B\,\rangle$
standing for an average over the nucleons' positions
$x_i$ and $y_j$ in the transverse plane. Each pair
$\{i,j\}$ enters the product only once, meaning that
each nucleon from the projectile nucleus can scatter
on each nucleon on the target once but no more than once.

The total interaction cross section is
\begin{equation}
\label{stot}
\sigma_{AB}^{tot}\,=\,\frac{4\pi}{k}
\mathrm{Im}F_{AB}^{el}(q=0)\,=\,
\int d^{\,2}b\,
\bigl[1\,-\,S_{AB}(b)\bigr]
\end{equation}
while the integrated elastic cross section is
\begin{equation}
\label{intsel}
\sigma_{AB}^{el}\,=\,\int d^{\,2}b\,
\bigl[1\,-\,S_{AB}(b)\bigr]^2.
\end{equation}
The difference between these two values determines
the total inelastic, or reaction, cross section,
\begin{equation}
\label{sr}
\sigma_{AB}^{r}\,=\,\sigma_{AB}^{tot}
-\sigma_{AB}^{el}\,=\,
\int d^{\,2}b\,
\bigl[1\,-\,|S_{AB}(b)|^2\bigr].
\end{equation}

\section{Generating function}
A main obstacle to dealing with the Glauber
amplitude (\ref{I}) is its complicated combinatorial structure.
To treat it analytically we firstly rewrite it more
explicitly through nucleon distributions in the colliding
nuclei,
\begin{eqnarray}
\label{SAB}
S_{AB}(b)\,&=&\,\int \prod_{i=1}^A d^{\,2}x_i
\int \prod_{j=1}^B d^{\,2}y_j
\rho_A^\bot(x_1-b,\ldots,\,x_A-b)\,
\rho_B^\bot(y_1,\ldots,\,y_B) \nonumber \\
&&\times\,
\left\{\prod\limits_{i\,j}\bigl[1+g\Gamma_{NN}(x_i-y_j)\bigr]
\right\}.
\end{eqnarray}
Here the nucleon densities in the transverse plane,
$\rho_{A,B}^\bot$, are determined through
three-dimensional ones integrated over longitudinal
coordinates,
$$
\rho_N^\bot(x_1,\ldots,x_N)\,=\,
\int \prod_{i=1}^N d z_i\,
\rho_N(z_1,x_1,\ldots,z_N,x_N),~~~~
\int \prod_{i=1}^N d^{\,3}r_i\,\rho_N(r_1,\ldots,r_N)\,=\,1.
$$
For later convenience an extra parameter $g$
counting the number of interactions is introduced
in Eq.~(\ref{SAB}), really $g=-1$.
We also assume in what follows that the three-dimensional
nuclear densities are reduced to the product of one-nucleon
densities,
$$
\rho_N(r_1,\ldots,r_N)\,=\,\prod_{i=1}^N \rho_N(r_i),~~~~
\int d^3r \rho_N(r)=1,
$$
and consequently
\begin{equation}
\label{rhoprod}
\rho_N^\bot(x_1,\ldots,x_N)\,=\,\prod_{i=1}^N \rho_N^\bot(x_i),~~~~
\int d^{\,2}x \rho_N^\bot(x)=1.
\end{equation}
This assumption means the nucleon-nucleon correlations are neglected.

The next step is to represent Eq.~(\ref{SAB}) as a functional
integral. To this end let us consider the identity
\begin{eqnarray}
\label{GI}
&&C_0
\int D\Phi D\Phi^*\exp\biggl\{
-\int d^{\,2}x d^{\,2}y\,\Phi(x)\Delta^{-1}(x-y)\Phi^*(y)
\\
&&+\sum_i\Phi(x_i)+\sum_j\Phi^*(y_j)
\biggr\}\,
=\,\exp\biggl\{\sum\limits_{i,j}\Delta(x_i-y_j)
\biggr\}\,=\,
\prod\limits_{i,j}e^{\,\Delta(x_i-y_j)},\nonumber
\end{eqnarray}
where $C_0$ is the normalization constant and
the functional integral can be thought of as
an infinite product of two dimensional integrals
over the auxiliary fields $\Phi(x)$ at each space
point $x$.
The inverse of the propagator, $\Delta^{-1}(x-y)$,
is understood in a functional sense,
$\int d^{\,2}z\Delta^{-1}(x-z)\Delta(z-y)
=\delta^{(2)}(x-y)$.
If this function is chosen to obey
the equation
\begin{equation}
\label{Delta}
e^{\,\Delta(x-y)}\,-\,1\,=\,g\Gamma_{NN}(x-y),
\end{equation}
the right-hand side of Eq.~(\ref{GI}) recovers the product
in Eq.~(\ref{SAB}). The function $\Delta(x-y)$ plays a role
similar to that of the Mayer propagator (function) in statistical
mechanics, the analogy between Glauber theory and statistical
mechanics has been remarked on earlier (see, e.g., Ref.~\cite{Boreskov:1987yt}).
Then we get
\begin{eqnarray}
S_{AB}(b)\,&=&\,
C_0
\int D\Phi D\Phi^*\exp\left\{
-\int d^{\,2}x d^{\,2}y\,\Phi(x)\Delta^{-1}(x-y)\Phi^*(y)
\right\} \nonumber \\
&&\times
\left[\int d^{\,2}x\,\rho_A^\bot(x-b)e^{\Phi(x)}\right]^A
\left[\int d^{\,2}y\,\rho_B^\bot(y)e^{\Phi^*(y)}\right]^B.
\end{eqnarray}
This form suggests that it is natural
to introduce the generating function,
\begin{eqnarray}
\label{Z}
Z(u,v)\,&=&\,
\int D\Phi D\Phi^*\exp\left\{
-\int\! d^{\,2}x d^{\,2}y\,\Phi(x)\Delta^{-1}(x-y)\Phi^*(y)
\right. \\
&&\,\left.+u\int d^{\,2}x\,\rho_A^\bot(x-b)e^{\Phi(x)} +v\int\!
d^{\,2}x\,\rho_B^\bot(x)e^{\Phi^*(x)} \right\},\nonumber
\end{eqnarray}
so that
\begin{equation}
\label{ddZ}
S_{AB}(b)\,=\,\frac 1{Z(0,0)}
\frac{\partial^A}{\partial u^A}
\frac{\partial^B}{\partial v^B}\,
Z(u,v)\biggl|_{u=v=0}.
\end{equation}

The generating function (\ref{Z}) is the focus of the present paper.
Being analytically evaluated it comprises all interaction orders
of the Glauber theory for nucleus-nucleus collision.

\section{Explicit evaluation of the generating function}
As it has been mentioned above the Glauber theory essentially
relies on the short-distance nature of the nucleon-nucleon
interaction.
The same property, the small interaction range, makes
the complex functional integral (\ref{Z}) feasible.
The standard parametrization
of the elastic nucleon-nucleon scattering amplitude is
\begin{equation}
\label{f_NN}
f_{NN}^{el}(q)\,=\,ik\frac{\sigma_{NN}^{tot}}{4\pi}
e^{-\frac 12\beta q^2},
\end{equation}
where $\sigma_{NN}^{tot}$ is the total nucleon-nucleon
cross section. It gives according to Eq.~(\ref{chi})
\begin{equation}
\label{GNN}
\Gamma_{NN}(x)\,=\,\frac{\sigma_{NN}^{tot}}{4\pi\beta}
e^{-\frac {x^2}{2\beta}},
\end{equation}
the value $a = \sqrt{2\pi\beta}$
being of the order of the interaction radius.
Assuming $a$ to be small at the nuclear scale
the nucleon-nucleon amplitude can be treated
as a point-like function,
\begin{equation}
\label{Gdelta}
\Gamma_{NN}(x)\,\simeq\,\frac 12\,\sigma_{NN}^{tot}\,
\delta^{(2)}(x).
\end{equation}
If $\Delta(x-y)$ is point-like the integrals
over the $\Phi(x)$ fields in Eq.~(\ref{Z}) are independent for
different coordinate values. This turns the functional integral
into the infinite product of finite-dimension integrals,
which can be separately evaluated for each $x$.

To do this accurately we replace the continuous integrals
in the exponent by discretized sums. The discrete version
of the identity (\ref{GI}) reads
\begin{eqnarray}
&&C_0\prod\limits_{x_n}\int \frac{d\Phi(x_n) d\Phi^*(x_n)}{2\pi}
\exp\left\{-\sum\limits_{n}
\frac 1y\Phi(x_n)\Phi^*(x_n)
+\sum_i\Phi(x_i)+\sum_j\Phi^*(y_j)
\right\}\nonumber \\
&&\,=\,
\label{GIdis}
\exp\bigl\{y\,\sum\limits_{i,j}\delta_{x_i,y_j}\bigr\},
\end{eqnarray}
where $\delta_{x_i,y_j}$ is Kronecker symbol for the discrete
nucleons' coordinates,
$\delta_{x_i,y_j}/a^2\to \delta^{(2)}(x_i-y_j)$
for $a\to 0$.
Since
$$
e^{\,y\delta_{x_i,y_j}}\,=\,1\,+\,(e^{\,y}-1)\delta_{x_i,y_j}
$$
the right hand side of the identity (\ref{GIdis}) yields
$$
\prod\limits_{i,j}\left[1+g\frac 12 \frac{\sigma_{NN}^{tot}}{a^2}
\delta_{x_i,y_j}\right]\,\to\,
\prod\limits_{i,j}\left[1+g\Gamma_{NN}(x_i-y_j)\right],
$$
whereas Eq.~(\ref{Delta}) translates into
\begin{equation}
\label{yd}
e^{\,y}\,-\,1\,=\,g\frac 12 \frac{\sigma_{NN}^{tot}}{a^2}.
\end{equation}
The generating function becomes the product
of independent integrals at the points $x_i$,
\begin{eqnarray}
\label{ZPhi}
Z(u,v)\,&=&\,
\prod\limits_{x_i}\int \frac{d\Phi(x_i) d\Phi^*(x_i)}{2\pi}
\exp\bigl\{
-\,\frac 1{y}\,
\Phi(x_i)\Phi^*(x_i)  \nonumber \\
&&\, +\,u\,a^2\rho_A^\bot(x_i-b)e^{\Phi(x_i)}
+v\,a^2\,\rho_B^\bot(x_i)e^{\Phi^*(x_i)}
\bigr\},\nonumber
\end{eqnarray}
or, after evaluating $\Phi(x)$ integrals in (\ref{ZPhi}),
\begin{equation}
\label{MN}
Z(u,v)\,=\,
\exp\left\{
\sum\limits_{x_i}\ln\left(y
\sum\limits_{M,N\ge 0}\frac{e^{\,y\,M\,N}}{M!N!}
\bigl[a^2 u\rho_A^\bot(x_i-b)\bigr]^M
\bigl[a^2 v\rho_B^\bot(x_i)\bigr]^N
\right)
\right\}.
\end{equation}
The densities are slowly varying at the size $a$,
which allows to replace the sum over $x_i$ with the integral,
\begin{eqnarray}
\label{Zuv}
Z(u,v)\,&=&\,C\,e^{W_y(u,v)} \\
\label{Wy}
W_y(u,v)\,&=&\, \frac 1{a^2}\int d^{\,2}x\,
\ln\bigl(\!\!
\sum\limits_{M\le A,N\le B}
\frac{z_y^{M\, N}}{M!N!}
\bigl[a^2 u\rho_A^\bot(x-b)\bigr]^M
\bigl[a^2 v\rho_B^\bot(x)\bigr]^N
\bigr),
\end{eqnarray}
with $u$- and $v$-independent constant $C$
being irrelevant in Eq.~(\ref{ddZ}) and
$$
z_y\,=\,1+g\frac 12 \frac{\sigma_{NN}^{tot}}{a^2}.
$$
The sums over $M$
and $N$ can always be truncated up to $A$ and $B$
because the higher terms obviously do not
contribute to the derivatives in Eq.~(\ref{ddZ}).
Put differently, the number of contributions to the generating
function does not exceed the number of various brackets in
the initial product (\ref{I}).

\section{Relation to the known approximations}
The formulas (\ref{Zuv}) and (\ref{Wy}) are the net result for
the generating function. To elaborate it further
we expand $W_y(u,v)$ into the series built of the densities overlaps,
\begin{equation}
\label{tmn}
t_{m,n}(b)=
\frac 1{a^2}\int d^{\,2}x\,
\bigl[a^2\rho_A^\bot(x-b)\bigr]^m\,
\bigl[a^2\rho_B^\bot(x)\bigr]^n.
\end{equation}
Since $t_{0,1}(b)=t_{1,0}(b)=1$
we have $W_y(u,v)=u+v+F(u,v)$
and the amplitude reads
\begin{equation}
\label{Fuv}
S_{AB}(b)\,=\,\sum\limits_{k,j\le A,B}
\frac{A!B!}{(A-k)!(B-j)!}\,
\frac 1{k!}\frac{\partial^{\,k}}{\partial\, u^{k}}\,
\frac 1{j!}\frac{\partial^{\,j}}{\partial\, v^{j}}\,
e^{\,F(u,v)}\bigl|_{u=v=0}.
\end{equation}
For $A,B \gg 1$ one may assume that $k,j\ll A,B$
and
$A!/(A-k)!\cdot B!/(B-j)!\approx A^k \cdot B^j$, which gives
\begin{equation}
\label{FAB}
S_{AB}(b)\,\approx\,e^{\,F(A,B)}.
\end{equation}

Really the functions $t_{m,n}(b)$ decrease as the indices
$m$ and $n$ grow. Keeping only the lowest, $m=n=1$,
we arrive at the well-known optical approximation~\cite{Czyz:1969jg}
\begin{equation}
\label{t11}
F(A,B)\,=\,
-\frac 12 \,\frac{\sigma_{NN}^{tot}}{a^2}\,
T_{AB}(b),~~~~
T_{AB}(b)\,=\,A\, B\, t_{1,1}(b).
\end{equation}
The optical approximation is equivalent
to the requirement that each nucleon from one nucleus
interacts with another nucleus no more than once.

Another known approximation easily reproduced here
is the rigid target (or projectile)
approximation~\cite{Bialas:1977pd, Alkhazov:1977ur}.
It allows any nucleon from the projectile
to interact with several nucleons from the target,
whereas any target nucleon can interact
no more than once.
Though it seems to be rather natural
when the atomic weight of the projectile is significantly
smaller than that of the target nucleus,
this approximation works fairly good even
for the equal atomic weights
\cite{Braun:1988pk}.
It requires one density, say, $\rho_A^\bot(x)$,
to be kept in the formula (\ref{Zuv}) only in the linear
order, permitting at the same time any powers
of $\rho_B^\bot(x)$.
It gives
\begin{eqnarray}
W_y(u,v)\,&=&\, \frac 1{a^2}\int d^{\,2}x\,
\ln\bigl(
\sum\limits_{N}\frac 1{N!}\bigl[a^2 v\rho_B^\bot(x)\bigr]^N
+\bigl[a^2\rho_A^\bot(x-b)\bigr]
\sum\limits_{N}\frac{z_y^N}{N!}\bigl[a^2 v\rho_B^\bot(x)\bigr]^N
\bigr)\nonumber \\
&=&\,v\,+\,u\int d^{\,2}x\,\rho_A^\bot(x-b)
e^{a^2\,v\,(z_y-1)\rho_B^\bot(x)}, \nonumber
\end{eqnarray}
yielding the generating function
$$
Z(u,v)\,=\,e^{v + u\, T_{rg}(v,b)},~~~~
T_{rg}(v,b)\,=\,\sum_{n=0}^\infty \frac 1{n!}\,
t_{1,n}(b)\,[(z_y-1)v]^n,
$$
which produce for $B\gg 1$
\begin{equation}
\label{rt}
S_{AB}(b)\,=\,\bigl[T_{rg}(b)\bigr]^A,~~~~
T_{rg}(b)\,=\,\int d^{\,2}x\,\rho_A^\bot(x-b)\,
e^{-\frac 12 \sigma_{NN}^{tot}\rho_B^\bot(x)}.
\end{equation}

\section{Results for $^{12}$C -- $^{12}$C scattering}
Below we present the results obtained with
the full generating function (\ref{Zuv})
for the $^{12}$C -- $^{12}$C scattering
in the energy interval 800 -- 1000~MeV per projectile
nucleon, where the experimental data exist~\cite{Ozava}.
The total cross section $\sigma_{NN}^{tot}=43$~mb
has been taken from averaging over $pp$ and $pn$ values,
and the slope value has been chosen to be $\beta = 0.2$~fm$^2$.
The nucleon density is parametrized by
harmonic oscillator distribution well suited for light
nuclei with the atomic weight $A\le 20$,
\begin{equation}
\label{HO}
\rho_A(r)\,=\,\rho_0\bigl[1+\frac 16\,(A-4)\frac{r^2}{\lambda^2}
\bigr]e^{-\frac{r^2}{\lambda^2}},
\end{equation}
with $\rho_0$ being the normalization, and the factor $\lambda$
being adjusted to match the nuclear mean square radius,
$R_{\mathrm{rms}}=\sqrt{r_A^2}$, where
$r_A^2=\int d^3r\,r^2\rho_A(r)$.

Upon evaluating $W_y(u,v)$ through all
$t_{mn}(b)$ functions (\ref{tmn}) for $m,n \le A=12$
in the parametrization (\ref{HO})
with $R_{\mathrm{rms}}=2.49$~fm fitted for this parametrization
in Ref.~\cite{Merino:2009yj} from Monte-Carlo simulation
of a $^{12}$C -- $^{12}$C collision, we have calculated
the reaction cross section (\ref{sr}) and the total cross section
(\ref{stot}).
Table I compares their values obtained in the optical approximation
(\ref{t11}), in the rigid target approximation (\ref{rt})
and with the full generating function for two cases, first
assuming $A\gg 1$ and using the approximate formula (\ref{FAB}) and second
by exact differentiating the generating function.

\medskip
Table I. The reaction and the total cross sections of
the $^{12}$C -- $^{12}$C collision
at the energy 950~MeV per nucleon
and $R_{\mathrm{rms}}=2.49$~fm. The first two columns present
the results of the optical and rigid target approximations,
and the second two columns present the results obtained with
the full generating function, assuming $A\gg 1$
(third column) and exactly differentiating it
(fourth column).

\begin{center}
\begin{tabular}{c c c c c}\hline \hline
&Optical &
 Rigid target &
Assuming &
Exact \\
&~~approximation~~ &~~approximation~~&~~ $A\gg 1$~~ &~~differentiating~~\\
\hline
$\sigma^r$, mb & 952 & 911 &
857 & 867 \\
\hline
$\sigma^{\,tot}$, mb & 1572 &
1470 & 1371 & 1363 \\
\hline \hline
\end{tabular}
\end{center}

The last two numbers in the upper row of the table
are in reasonable
agreement with the experimental value
$853\pm 6$~mb~\cite{Ozava}. One should bear in mind
that the experimentally measured value actually refers
to the so-called interaction cross section rather than
to the reaction one. The difference between
them can be at the several percents level~\cite{Novikov:2013zdw}.
At the same time the obtained values are close to those of the Monte Carlo
calculations with the same parameters and the density parametrization
\cite{Merino:2009yj}.

A word of caution with respect to the formula (\ref{FAB})
is needed here.
When the generating function is exactly differentiated
in Eq.~(\ref{Fuv}), the terms with $m>A$ and $n>B$
in the function
$F(u,v)\,=\,\sum_{m,n}F_{m,n}u^m v^n$
do not evidently contribute.
The reliability of the approximation (\ref{FAB})
implies the series for the function
$F(A,B)$ to be truncated at $m=A$ and $n=B$.
Leaving more terms does not
improve, but may worsen the accuracy.
Of course, it does not apply to the optical (\ref{t11})
or rigid target (\ref{rt}) approximations,
where all the extra terms are already dropped out.

Representing the amplitude (\ref{SAB}) as a power
series in the parameter $g$
enables one to pick up the individual contribution
of $n$-fold interaction
as a coefficient in front of the $g^n$ term.
They are large when taken separately,
but due to the opposite signs they almost
cancel each other returning a final sum
much smaller than any of them.
Thus the treatment of the amplitude
in terms of the number of interactions seems to be
quite unreasonable.

\section{Halo nuclei scattering}
An interesting topic to apply our method to
is the scattering of halo nuclei. The distinguishing
feature of these nuclei is their large size
exceeding that of the stable nuclei.
They are assumed
to be a composite systems of a core and a halo
\cite{Tanihata:1995yv,Ershov:1997az},
the density distribution being the sum of
the two components
\cite{Alkhazov:2004syy, Ilieva:2012vhs,
Korsheninnikov:1997qta, Zhukov:1993aw},
\begin{equation}
\label{halo}
\rho(r)\,=\,N_c\rho_c(r)\,+\,N_v\rho_v(r).
\end{equation}
The first and the second terms stand here for the core
and for the halo densities, $N_c$ is the number of the nucleons
in the core, and $N_v$ is the number of the valence neutrons
in the halo \cite{Hassan:2015dfa}.
Both the densities are taken in the Gaussian form, and the core density reads
$$
\rho_c(r)\,=\,\frac 1{\pi^{\frac 32}a_c^3}\,e^{-\frac{r^2}{a_c^2}},
~~~~a_c=\sqrt{2/3}\,R_c,
$$
where $R_c$ is the core mean square radius,
while the second part, $\rho_v$, usually admits
three different parametrizations
depending on the shell state the halo
neutrons are supposed to occupy \cite{Hassan:2015dfa}:
\begin{equation}
\label{params}
\begin{array}{ccc}
  \rho_v^G(r)\,=\,\frac 1{\pi^{\frac 32}a_G^3}e^{-\frac{r^2}{a_G^2}},~~~&
  ~\rho_v^O(r)\,=\,\frac 2{3\pi^{\frac 32}a_O^5}\,r^2e^{-\frac{r^2}{a_O^2}},~~~&
  ~\rho_v^{2S}(r)\,=\,\frac 2{3\pi^{\frac 32}a_{2S}^3}
   \left(\frac{r^2}{a_{2S}^2}-\frac 32\right)^2
  e^{-\frac{r^2}{a_{2S}^2}},
 \\
 a_G=\sqrt{2/3}\,R_v,  & a_O=\sqrt{2/5}\,R_v, & a_{2S}=\sqrt{2/7}R_v,
\end{array}
\end{equation}
where $R_v$ is the halo mean square radius. There exists also
a "non halo" distribution that neglects the halo and uses
the density (\ref{HO}) fitted to the experimental matter radius
\cite{Korsheninnikov:1997qta}.

In Table II we present the reaction cross sections
for  $^{11}$Li -- $^{12}$C, $^{11}$Be -- $^{12}$C
and $^{14}$Be -- $^{12}$C scattering
at the energy 790~MeV per nucleon obtained
through exactly differentiated full generating function.
The calculations have been carried out with the density (\ref{halo})
(normalized to unity) for all three parameterizations
with~\cite{Hassan:2015dfa}

\begin{tabular}{lllll}
$R_c=2.50$~fm, & $R_v=5.04$~fm, & $N_c=9$ &,
$N_v=2$ &~for~ $^{11}$Li,\\
$R_c=2.30$~fm, & $R_v=5.39$~fm, & $N_c=10$ &,
$N_v=1$ &~for~ $^{11}$Be,\\
$R_c=2.59$~fm, & $R_v=5.45$~fm, & $N_c=12$ &,
$N_v=2$ &~for~ $^{14}$Be.
\end{tabular}

\noindent
Besides we add the results for
the "non halo" distributions with $R_{\mathrm{rms}}=3.12$~fm
for $^{11}$Li, $R_{\mathrm{rms}}=2.73$~fm for $^{11}$Be
and $R_{\mathrm{rms}}=3.16$~fm for $^{14}$Be
\cite{Ozava}. The parametrization of $^{12}$C is the same as in Table~I.

\medskip
Table II. The reaction cross sections in mb for
$^{11}$Li -- $^{12}$C, $^{11}$Be -- $^{12}$C
and $^{14}$Be -- $^{12}$C collisions
at the energy 790~MeV per nucleon. The first three columns
are for the three density parametrizations (\ref{params}),
and the fourth column is for the "non halo" distributions. The experimental
points in the fifth column are taken from Ref.~\cite{Ozava}.

\begin{center}
\begin{tabular}{ c c c c c c}\hline \hline
&~~~~$G$~~~~& ~~~~$O$~~~~& ~~~~$2S$~~~~ &"Non halo"&Expt\\
\hline
$^{11}$Li~~ & 1024 & 1031 &
1033 & 1021 & 1040 $\pm$ 60 \\
\hline
$^{11}$Be~~ & 911 &
918 & 916 & 914 & 942 $\pm$ 8 \\
\hline
$^{14}$Be~~ & 1120 &
1128 & 1131 & 1103 & 1139 $\pm$ 90 \\
\hline \hline
\end{tabular}
\end{center}

The results for $^{11}$Li and $^{14}$Be are in a quite good
agreement with the experimental data especially regarding
the relatively large error bars. The evaluated cross sections
for the $^{11}$Be beam are systematically smaller although
the deviation is not large.
One might doubt whether the parameters
are determined for this case with the proper accuracy.
On the other hand the reason for the discrepancy could be
in the nucleons' correlations.

It is worth pointing out that it is the reaction cross sections
that have been calculated here, whereas
the experimentally measured value is the interaction cross section.
It is smaller than the reaction one, but their difference does not exceed
1 -- 2\% \cite{Novikov:2013zdw}.

\section{Conclusion}
In this article we set up a novel
approach to deal with Glauber
amplitudes for nucleus-nucleus scattering at energies
higher than several hundreds~of~MeV per nucleon. It is based on the closed
expression obtained for the generating function. The main advantage
it has is in a relatively simple analytical form that allows one
to carry out calculations avoiding
the complexities encountered in the Monte Carlo technique.
As an example we apply our method to $^{12}$C -- $^{12}$C
scattering at the energy 950~MeV per nucleon
for which there exist the experimental data~\cite{Ozava}.
We have calculated the reaction and the total cross sections
for the mean square nuclear
radius $R_{\mathrm{rms}}=2.49$~fm, the value
taken from the Monte Carlo analysis
in the harmonic oscillator parametrization in
Ref.~\cite{Merino:2009yj}. Our results are in good
agreement with those obtained in that paper.
As another example we have calculated
the cross section of several halo nuclei scattering
on a $^{12}$C target at the energy 790~Mev per nucleon.

The proposed generating function (\ref{Zuv}) is appropriate
for any pairs of colliding nuclei regardless their atomic
weight. Apart from the above-considered integrated cross sections
it can provide a consistent evaluation of the differential
elastic cross section (\ref{dcs}) as well.
In this case, however, one has to account for the Coulomb
corrections at small scattering angles for heavy nuclei

Taking the nucleon density as the product
of single particle ones (\ref{rhoprod}) we thereby neglect
the nucleon-nucleon correlations. The particular correlations
can be, in principle, accounted for in our approach, provided
an appropriate wavefunction is known.

We are grateful to I.S. Novikov for helpful discussions.


\begin{thebibliography}{**}
%\cite{Tanihata:1985zq}
\bibitem{Tanihata:1985zq}
I.~Tanihata, H.~Hamagaki, O.~Hashimoto, S.~Nagamiya,
Y.~Shida, N.~Yoshikawa, O.~Yamakawa, K.~Sugimoto,
T.~Kobayashi and D.~E.~Greiner,
\textit{et al.}
{\em Measurements of Interaction Cross-Sections and Radii
of He Isotopes},
Phys. Lett. B \textbf{160}, 380 (1985).
% doi:10.1016/0370-2693(85)90005-X

%\cite{Tanihata:1986kh}
\bibitem{Tanihata:1986kh}
I.~Tanihata, H.~Hamagaki, O.~Hashimoto, Y.~Shida,
N.~Yoshikawa, K.~Sugimoto, O.~Yamakawa, T.~Kobayashi
and N.~Takahashi,
{\em Measurements of Interaction Cross-Sections
and Nuclear Radii in the Light p-Shell Region},
Phys. Rev. Lett. \textbf{55}, 2676 (1985).
% doi:10.1103/PhysRevLett.55.2676

%\cite{Smith:2008zh}
\bibitem{Smith:2008zh}
M.~Smith, M.~Brodeur, T.~Brunner, S.~Ettenauer, A.~Lapierre,
R.~Ringle, V.~L.~Ryjkov, F.~Ames, P.~Bricault,
G.~W.~F.~Drake, \textit{et al.}
{\em First Penning-trap Mass Measurement in the Millisecond
Half-Life Range: The Exotic Halo Nucleus Li-11},
Phys. Rev. Lett. \textbf{101}, 202501 (2008).
% doi:10.1103/PhysRevLett.101.202501
% [arXiv:0807.1260 [nucl-ex]].
%113 citations counted in INSPIRE as of 22 Jan 2021

%\cite{Bertulani:2009zw}
\bibitem{Bertulani:2009zw}
C.~A.~Bertulani,
{\em Nuclear Reactions},
arXiv:0908.3275.

\bibitem{Glaub1}
R.J. Glauber, {\em Cross-sections in deuterium at high-energies}
Phys. Rev. \textbf{100}, 242 (1955).


%\cite{Glauber:1970jm}
\bibitem{Glauber:1970jm}
R.~J.~Glauber and G.~Matthiae,
{\em High-energy scattering of protons by nuclei},
Nucl. Phys. B \textbf{21}, 135 (1970).
% doi:10.1016/0550-3213(70)90511-0

%\cite{Czyz:1969jg}
\bibitem{Czyz:1969jg}
W.~Czyz and L.~C.~Maximon,
{\em High-energy, small angle elastic scattering
of strongly interacting composite particles},
Annals Phys. \textbf{52}, 59 (1969).
% doi:10.1016/0003-4916(69)90321-2

%\cite{Bialas:1977pd}
\bibitem{Bialas:1977pd}
A.~Bialas, M.~Bleszynski and W.~Czyz,
{\em Relation Between the Glauber Model
and Classical Probability Calculus},
Acta Phys. Pol. B \textbf{8}, 389 (1977).

%\cite{Alkhazov:1977ur}
\bibitem{Alkhazov:1977ur}
G.~D.~Alkhazov, T.~Bauer, R.~Bertini, L.~Bimbot,
O.~Bing, A.~Boudard, G.~Bruge,
H.~Catz, A.~Chaumeaux, P.~Couvert, \textit{et al.}
{\em Elastic and Inelastic Scattering of 1.37-GeV alpha Particles
from Ca-40, Ca-42, Ca-44, Ca-48},
Nucl. Phys. A \textbf{280}, 365 (1977).
% doi:10.1016/0375-9474(77)90611-X

%\cite{Zadorozhnyi:1983fz}
\bibitem{Zadorozhnyi:1983fz}
A.~M.~Zadorozhnyi, V.~V.~Uzhinsky and S.~Y.~Shmakov,
{\em A Stochastic Method of Calculating Nucleus-nucleus
Scattering Characteristics in the Eikonal Approach},
Sov. J. Nucl. Phys. \textbf{39}, (1984), 729 JINR-P2-83-544.

\bibitem{Gareev}
F.~A.~Gareev, S.~N.~Ershov, G.~S.~Kazacha, E.~F.~Svinareva,
S.~Y.~Shmakov and V.~V.~Uzhinski,
{\em Study of properties of exotic nuclei using elastic scattering.
Theoretical consideration},
F.A. Gareev et al., Yad. Fiz. \textbf{58}, 620 (1995)
Phys. At. Nucl. \textbf{58}, 564 (1995).

\bibitem{Lobodenko}
G.D. Alkhazov and A.A. Lobodenko,
{\em Reaction cross sections for collisions involving
exotic light nuclei within the glauber approach},
Yad. Fiz. \textbf{70}, 98 (2007)
Phys. At. Nucl. \textbf{70}, 93 (2007).


%\cite{Merino:2009yj}
\bibitem{Merino:2009yj}
C.~Merino, I.~S.~Novikov and Y.~M.~Shabelski,
{\em Nuclear Radii Calculations in Various Theoretical
Approaches for Nucleus-Nucleus Interactions},
Phys. Rev. C \textbf{80}, 064616 (2009).
% doi:10.1103/PhysRevC.80.064616
% [arXiv:0907.1697 [nucl-th]].
% 8 citations counted in INSPIRE as of 12 Mar 2021

% %\cite{Loizides:2014vua}
% \bibitem{Loizides:2014vua}
% C.~Loizides, J.~Nagle and P.~Steinberg,
% {\em Emproved version of the PHOBOS Glauber Monte Carlo},
% SoftwareX \textbf{1-2} (2015), 13-18
% % doi:10.1016/j.softx.2015.05.001
% [arXiv:1408.2549 [nucl-ex]].


%\cite{Pajares:1983gw}
\bibitem{Pajares:1983gw}
C.~Pajares and A.~V.~Ramallo,
{\em Effects of the multiple scattering structure
in the propagation of hadronic properties
in nucleus-nucleus collisions},
Phys. Rev. D \textbf{31}, 2800 (1985).
% doi:10.1103/PhysRevD.31.2800
% 45 citations counted in INSPIRE as of 26 Jan 2021

%\cite{Braun:1988pk}
\bibitem{Braun:1988pk}
V.~M.~Braun and Y.~M.~Shabelski,
{\em Multiple Scattering Theory for Inelastic Processes},
Int. J. Mod. Phys. A \textbf{3}, 2417 (1988).
% doi:10.1142/S0217751X8800103X
% 28 citations counted in INSPIRE as of 26 Jan 2021

%\cite{Boreskov:1987yt}
\bibitem{Boreskov:1987yt}
K.~G.~Boreskov and A.~B.~Kaidalov,
{\em Nucleus-nucleus Scattering in the Glauber Approach},
Sov. J. Nucl. Phys. \textbf{48}, 367 (1988).
%31 citations counted in INSPIRE as of 30 Jan 2021

\bibitem{Ozava}
A. Ozawa, T. Suzuki, and I. Tanihata,
{\em Nuclear size and related topics},
Nucl. Phys.A693, 32 (2001).

%\cite{Novikov:2013zdw}
\bibitem{Novikov:2013zdw}
I.~S.~Novikov and Y.~Shabelski,
{\em Complete Glauber calculations of reaction
and interaction cross sections for light-ion collisions},
Phys. At. Nucl. \textbf{78}, 951 (2015).
% doi:10.1134/S106377881507011X
[arXiv:1302.3930 [nucl-th]].

%\cite{Tanihata:1995yv}
\bibitem{Tanihata:1995yv}
I.~Tanihata,
{\em Neutron halo nuclei},
J. Phys. G: Nucl. Part. Phys. \textbf{22}, 157 (1996).
% doi:10.1088/0954-3899/22/2/004
% 362 citations counted in INSPIRE as of 17 Oct 2021

%\cite{Ershov:1997az}
\bibitem{Ershov:1997az}
S.N. Ershov, T. Rogde, B.V. Danilin, J.S. Vaagen,I.J. Thompson,
F.A. Gareev,
{\em Halo excitation of He-6 in inelastic
and charge exchange reactions},
Phys. Rev. C \textbf{56}, 1483 (1997).
% doi:10.1103/PhysRevC.56.1483
% [arXiv:nucl-th/9705002 [nucl-th]].
% 45 citations counted in INSPIRE as of 17 Oct 2021

%\cite{Alkhazov:2004syy}
\bibitem{Alkhazov:2004syy}
G.~D.~Alkhazov, A.~V.~Dobrovolsky and A.~A.~Lobodenko,
{\em Matter density distributions and radii of light exotic nuclei
from intermediate-energy proton elastic scattering
and from interaction cross sections},
Nucl. Phys. A \textbf{734}, 361 (2004).
% doi:10.1016/j.nuclphysa.2004.01.066
% 11 citations counted in INSPIRE as of 17 Oct 2021

%\cite{Ilieva:2012vhs}
\bibitem{Ilieva:2012vhs}
S.~Ilieva, F.~Aksouh, G.~D.~Alkhazov, L.~Chulkov,
A.~V.~Dobrovolsky, P.~Egelhof, H.~Geissel, M.~Gorska,
A.~Inglessi, R.~Kanungo, \textit{et al.}
{\em Nuclear-matter density distribution in the neutron-rich
nuclei 12,14 Be from proton elastic scattering
in inverse kinematics},
Nucl. Phys. A \textbf{875}, 8 (2012).
% doi:10.1016/j.nuclphysa.2011.11.010
% 51 citations counted in INSPIRE as of 17 Oct 2021

%\cite{Korsheninnikov:1997qta}
\bibitem{Korsheninnikov:1997qta}
A.~A.~Korsheninnikov, E.~Y.~Nikolskii, C.~A.~Bertulani,
S.~Fukuda, T.~Kobayashi, E.~A.~Kuz\-min, S.~Mo\-mo\-ta,
B.~G.~Novatskii, A.~A.~Ogloblin, A.~Ozawa, \textit{et al.}
{\em Scattering of radioactive nuclei 6 He and 3 H by protons:
Effects of neutron skin and halo in 6 He, 8 He, and 11 Li},
Nucl. Phys. A \textbf{617}, 45 (1997).
% doi:10.1016/S0375-9474(96)00492-7
% 94 citations counted in INSPIRE as of 17 Oct 2021

%\cite{Zhukov:1993aw}
\bibitem{Zhukov:1993aw}
M.~V.~Zhukov, B.~V.~Danilin, D.~V.~Fedorov, J.~M.~Bang,
I.~J.~Thompson and J.~S.~Vaagen,
{\em Bound state properties of Borromean Halo nuclei:
He-6 and Li-11},
Phys. Rept. \textbf{231}, 151 (1993).
% doi:10.1016/0370-1573(93)90141-Y
% 633 citations counted in INSPIRE as of 17 Oct 2021

%\cite{Hassan:2015dfa}
\bibitem{Hassan:2015dfa}
M.~A.~M.~Hassan, M.~S.~M.~Nour El-Din, A.~Ellithi,
E.~Ismail and H.~Hosny,
{\em The effect of halo nuclear density on reaction
cross-section for light ion collision},
Int. J. Mod. Phys. E \textbf{24}, 1550062 (2015).
% doi:10.1142/S0218301315500627
% 1 citations counted in INSPIRE as of 17 Oct 2021


\end{thebibliography}
\end{document}